\documentclass[aps,prl,superscriptaddress,showpacs,twocolumn]{revtex4-1}
\usepackage{amsmath}
\usepackage{amssymb}
\usepackage{graphicx}
\usepackage{color}
\definecolor{orange}{rgb}{1,0.5,0}
\def\nm{{\ {\rm nm}}}						

\def\Hz{{\ {\rm Hz}}}						

\def\kr{{{k_R}}}							
\def\El{{{E_L}}}							
\def\kl{{{k_L}}}							
\def\Rb87{^{87}\rm{Rb}}					

\def\ex{{\mathbf e}_x}                            
\def\ey{{\mathbf e}_y}                            
\def\ez{{\mathbf e}_z}                            
\DeclareMathAlphabet\mathbfcal{OMS}{cmsy}{b}{n}

\newcommand{\ket}[1]{|{#1}\rangle}

\newcommand{\beq}{\begin{equation}}
\newcommand{\eeq}{\end{equation}}

\begin{document}
 
\title{Synthetic gauge fields in synthetic dimensions}

\author{A. Celi}

\affiliation{ICFO -- Institut de Ci\`encies Fot\`oniques,
Mediterranean Technology Park, E-08860 Castelldefels (Barcelona), Spain}

\author{P. Massignan }

\affiliation{ICFO -- Institut de Ci\`encies Fot\`oniques,
Mediterranean Technology Park, E-08860 Castelldefels (Barcelona), Spain}

\author{J. Ruseckas }

\affiliation{Institute of Theoretical Physics and Astronomy, Vilnius University,
A. Go\v{s}tauto 12, Vilnius 01108, Lithuania}

\author{N. Goldman}

\affiliation{Center for Nonlinear Phenomena and Complex Systems - Universit\'e Libre de Bruxelles, 231, Campus Plaine, B-1050 Brussels, Belgium}

\author{I. B. Spielman}

\affiliation{Joint Quantum Institute, University of Maryland, College Park, Maryland
20742-4111, USA}

\affiliation{National Institute of Standards and Technology, Gaithersburg, Maryland
20899, USA}

\author{G. Juzeli\=unas }

\affiliation{Institute of Theoretical Physics and Astronomy, Vilnius University,
A. Go\v{s}tauto 12, Vilnius 01108, Lithuania}

\author{M. Lewenstein }

\affiliation{ICFO -- Institut de Ci\`encies Fot\`oniques,
Mediterranean Technology Park, E-08860 Castelldefels (Barcelona), Spain}

\affiliation{ICREA -- Instituci\'{o} Catalana de Recerca i Estudis Avan\c{c}ats, E-08010 Barcelona, Spain}
\date{\today}

\begin{abstract}

We describe a simple technique for generating a cold-atom lattice pierced by a uniform magnetic field. Our method is to extend a one-dimensional optical lattice  into the ``dimension'' provided by the internal atomic degrees of freedom, yielding a synthetic 2D lattice. 
Suitable laser-coupling between these internal states leads to a uniform magnetic flux within the 2D lattice. We show that this setup reproduces the main features of magnetic lattice systems, such as the fractal Hofstadter butterfly spectrum and the chiral edge states of the associated Chern insulating phases.
\end{abstract}


\maketitle
Intense effort is currently devoted to the creation of gauge fields for electrically neutral
atoms \cite{Lewenstein2007,Bloch2008a,Dalibard2011,Goldman2013RPP}.  Following a number of theoretical \mbox{proposals} in presence  \cite{Ruostekoski:2002,Jaksch2003,Mueller2004,Sorosen2005,Eckart2005,Osterloh2005,Gerbier2010,Kitagawa2010,Kolovsky2011} or in absence of optical lattices \cite{Dum1996,Visser1998,Juzeliunas2004,Ruseckas2005,Juzeliunas2006,Spielman2009,Gunter2009}, synthetic
magnetic fields have been engineered both in vacuum~\cite{Lin2009b,Lin2009a,Spielman-Hall-effect,Wang2012,Cheuk2012} and in periodic lattices~\cite{Bloch2011,Struck2013,Bloch2013,Ketterle2013}.  
The addition of a lattice offers the advantage to engineer extraordinarily large magnetic fluxes, typically of the order of one magnetic flux quantum per plaquette~\cite{Ruostekoski:2002,Jaksch2003,Mueller2004,Osterloh2005,Gerbier2010}, which are out of reach using real magnetic fields in solid-state systems (e.g. artificial magnetic fields recently reported in graphene \cite{Columbia2013,Manchester2013,MIT2013}). Such cold-atom lattice configurations will  enable one to access striking properties, such as Hofstadter-like fractal spectra \cite{Hofstadter1976} and Chern insulating phases, in a controllable manner. Existing schemes for creating uniform magnetic fluxes
require several laser fields
and/or additional ingredients, such as tilted potentials~\cite{Jaksch2003,Osterloh2005},
superlattices~\cite{Gerbier2010}, or lattice-shaking methods~\cite{Eckart2005,Zenesini2009,Eckart2010,Kolovsky2011,Struck2011,Hauke2012}. 
Experimentally, strong staggered magnetic flux configurations have been reported~\cite{Bloch2011,Struck2013}, and very recently also uniform ones \cite{Bloch2013,Ketterle2013}.  Besides, an alternative route is offered by optical flux lattices  \cite{Dudarev2004,Cooper2011a,Cooper2011,Juz-Spielm2012}.

\begin{figure}
\begin{centering}
\includegraphics[width=\columnwidth]{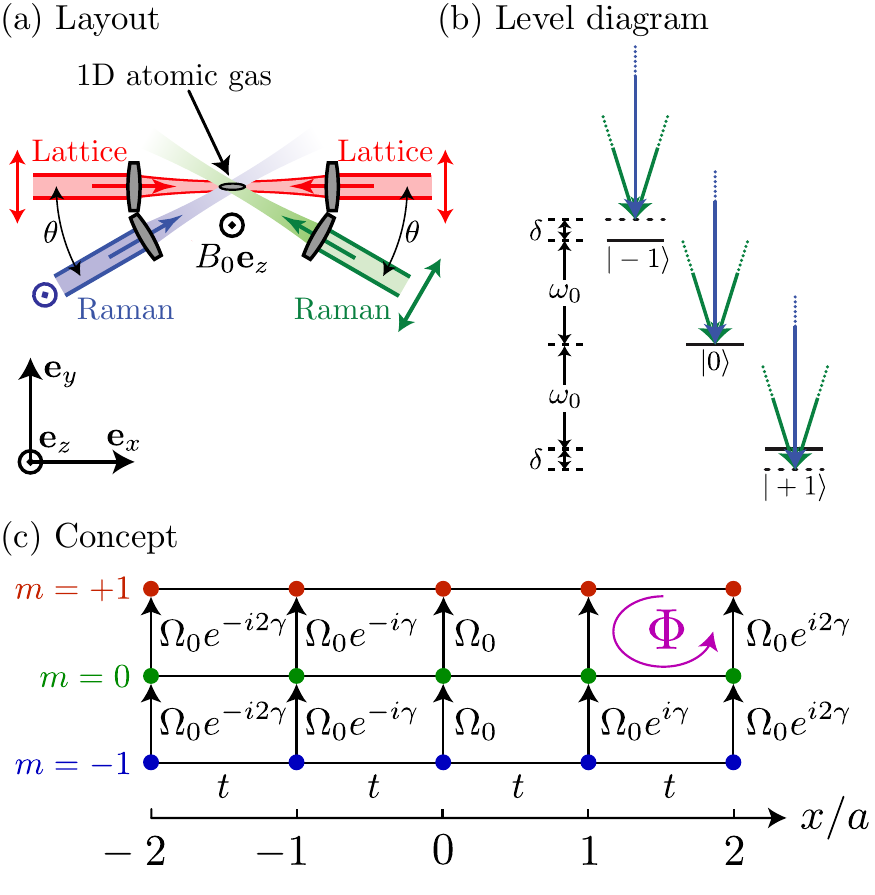}
\end{centering}
\caption{\label{fig:1} (a) Proposed experimental layout with $\Rb87$.  A pair of counter-propagating $\lambda=1064\nm$ lasers provide a $5\El$ deep optical lattice lattice with period $a=\lambda/2$.  A pair of ``Raman'' laser beams with wavelength $\lambda_R = 790\nm$, at angles $\pm\theta$ from $\ex$, couple the internal atomic states with recoil wavevector $\kr = 2\pi\cos(\theta)/\lambda_R$.  The laser beams' polarizations -- all linear -- are marked by symbols at their ends.  (b) Raman  couplings in the $F=1$ manifold.  The transitions are induced by the 
 beams depicted in (a). (c) Synthetic 2D lattice with magnetic flux $\Phi=\gamma/2 \pi$ per plaquette ($\gamma=2\kr a$). Here $n=x/a$ ($m$)
labels the sites along $\ex$ (Zeeman sublevels). }
\end{figure}

In all of these lattice schemes, the sites are identified by their location in space.  This need not be the case: the available spatial degrees of freedom can be augmented by employing the internal atomic ``spin'' degrees of freedom as an extra, or synthetic, lattice-dimension~\cite{Boada:2012}.  Here we demonstrate that this extra dimension can support a uniform magnetic flux, and we propose a specific scheme using a 1D optical lattice along with Raman transitions within the atomic ground state manifold (Fig. \ref{fig:1}).  The flux is produced by a combination of ordinary tunneling in real space and laser-assisted tunneling in the extra dimension creating the necessary Peierls phases. 
  Our proposal therefore extends the toolbox of existing techniques to create gauge potentials for cold atoms.
  
The proposed scheme distinguished by the naturally sharp boundaries in the extra dimension, a feature which greatly simplifies the detection of chiral edge states resulting from the synthetic magnetic flux~\cite{Goldman2010a,Stanescu:2010,Goldman:2012prl,Buchhold:2012,Goldman:2013PNAS}.  We demonstrate that the chiral motion of these topological edge states can be directly visualized using {\it in situ} images of the cloud, and we explicitly show their robustness against impurity scattering.  We also show that by using additional Raman and radio frequency transitions one can connect the edges in the extra dimension, providing a remarkably simple way to realize the
 fractal Hofstadter butterfly spectrum~\cite{Hofstadter1976}. 

{\it Model.} For specificity, consider $^{87}{\rm Rb}$'s $F=1$ ground state hyperfine manifold
~\cite{Note1}, composed of three magnetic sublevels $m_F= 0,\pm 1$, illuminated by the combination of optical lattice and Raman laser beams depicted in Fig. \ref{fig:1}(a) (additional lattice potentials along $\ey$ and $\ez$, confining motion to $\ex$ are not shown; ${\bf e}_{xyz}$ are the three Cartesian unit vectors).  In the schematic, the counter-propagating $\lambda=1064\nm$  lasers beams define the lattice with period $a=\lambda/2$, recoil momentum $\kl = 2\pi/\lambda$, and energy $\El = \hbar^2\kl^2/2m$ (where $m$ is the atomic mass).  We consider a sufficiently deep lattice $V_{\rm lat} = 5\El$ for the tight binding approximation to be valid, but shallow enough to avoid Mott-insulator physics.  For these parameters, the tunneling amplitude is $t = 0.065\El = h\times 133\Hz$.  The Raman lasers at wavelength $\lambda_R\approx790\nm$ intersect with opening angle $\theta$, giving an associated Raman recoil momentum $\kr = 2\pi\cos(\theta)/\lambda_R$.  The  Raman couplings recently exploited in experiment~\cite{Lin2009a,Lin2009b}, between the three magnetic sublevels $m_F=0,\pm 1$ of the $F=1$ ground-state manifold of $^{87}$Rb are shown in Fig. \ref{fig:1}(b).   The Raman transitions provide the hopping in the synthetic dimension  which require a minimum amount of laser light (less than 1\% required for existing schemes~\cite{Spielman2009}), minimizing spontaneous emission.  In addition, periodic boundary conditions in the synthetic direction can be created by coupling $m_F=+1$ to $m_F=-1$ using an off-resonant Raman transition from $\ket{F=1,m_F=+1}$ to an ancillary state, e.g., $\ket{F=2,m_F=0}$ (detuned by $\delta_{\rm pbc}$ and coupled with strength $\Omega_{R,{\rm pbc}}$), completed by a radio-frequency transition to $\ket{F=1,m_F=1}$ with strength $\Omega_{RF}$, giving a $\Lambda$-like scheme with strength $\Omega_{\rm pbc} = -\Omega_{R,{\rm pbc}} \Omega_{RF} / 2 \delta_{\rm pbc}$.

A constant magnetic field $B_0 \ez$ Zeeman splits the magnetic sublevels $\ket{m_F = \pm 1}$ by $\mp\hbar\omega_{0} = g_F \mu_{\rm B} B_0$, where $g_F$ is the Land\'e $g$-factor and $\mu_{\rm B}$ is the Bohr magneton, see Fig.~\ref{fig:1}(ab). 
 The Raman spin-flip transitions, detuned by $\delta$ from two-photon resonance, impart a $2\kr$ recoil momentum along $\ex$.   Taking $\hbar=1$, the laser fields can be described via a spatially periodic effective magnetic field 
\begin{equation}
{\boldsymbol{\Omega}}_{T}=\delta\mathbf{e}_{z}+\Omega_R\left[\cos\left(2\kr x\right)\mathbf{e}_{x}-\sin\left(2 \kr x\right)\mathbf{e}_{y}\right]\,,\label{eq:Omega}
\end{equation}
 which couples the hyperfine ground-states giving the  effective atom-light Hamiltonian~\cite{Goldman2013RPP,Dudarev2004,Deutsch1998,Juz-Spielm2012}
\begin{equation}
H_{\rm al}={\boldsymbol{\Omega}}_{T}\cdot\mathbf{F}=\delta F_{z}+(F_{+}e^{i\kr x}+F_{-}e^{-i\kr x})\Omega_R/2 \,,\label{eq:V}
\end{equation}
where the
operators \mbox{$F_{\pm}=F_{x}\pm i F_{y}$} act as $F_{+}\left|m\right\rangle =g_{F,m}\left|m+1\right\rangle $
with $g_{F,m}=\sqrt{F\left(F+1\right)-m\left(m+1\right)}$. Thus the
Raman beams sequentially couple states $m=-F,\,\ldots\,,F$, with each transition accompanied by an $x$-dependent phase.  This naturally generates Peierls phases for ``motion'' along the $m$ (spin) direction, denoted as ${\bf e}_m$.

The combination of the optical lattice along $\ex$ and the Raman-induced hopping along ${\bf e}_m$ yield an effective 2D lattice with one physical and one synthetic dimension, as depicted in Fig. \ref{fig:1}(c) for $F=1$.  For a system of length $L_x$ along $\ex$, the lattice has $N=L_x/a$ sites along $\ex$, and a width of $W=2F+1$ sites along ${\bf e}_m$. For $\delta=0$ the system is described by the  Hamiltonian
\begin{equation}
H=\sum_{n,m}\left(-ta_{n+1,m}^{\dag}+\Omega_{m-1}e^{-i\gamma n}a_{n,m-1}^{\dag}\right)a_{n,m}+{\rm H.c.}\,,
\label{eq:H}
\end{equation}
where $n$ labels the spatial index and $m$ labels the spin index; $\gamma=2\kr a$ sets the magnetic flux; $\Omega_m=\Omega_R g_{F,m}/2$ is the synthetic tunneling strength; and $a_{n,\, m}^{\dag}$ is the atomic creation operator in the dimensionally extended lattice.  This two-dimensional lattice is pierced by a uniform synthetic magnetic
flux $\Phi=\gamma/2\pi=\kr a/\pi$ per plaquette (in units of the Dirac flux quantum).
 The quantity $g_{F,m}$ is independent of $m$ for $F=1/2$ and $F=1$, but for larger $F$ hopping along ${\bf e}_m$ is generally non-uniform. 
 
\begin{figure}
\begin{centering}
\includegraphics[width=.98\columnwidth]{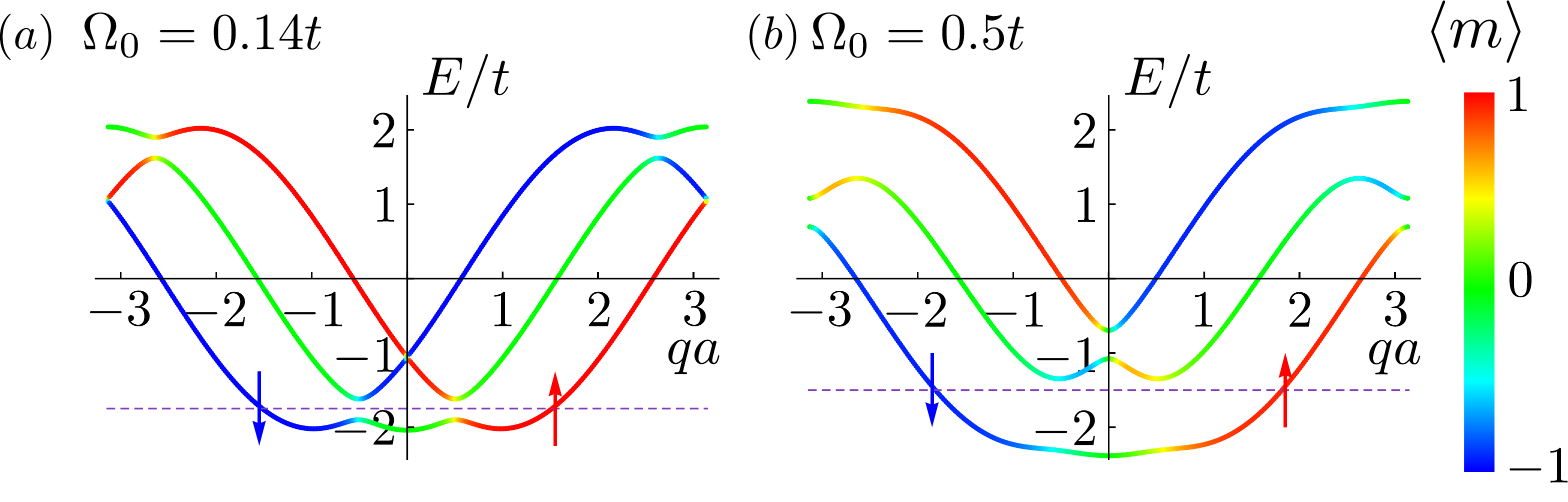}
\end{centering}
\caption{Spectrum for open boundary conditions: $F=1$, and $\Phi=\gamma/2\pi=1/2\pi$ flux per plaquette.  Colors specify the spin state $m$, as indicated. The ground state branch displays ``edges" corresponding to $m=\pm 1$.
}
\label{fig:openBCspectrum}
\end{figure}

{\it Open boundaries.}  Since $\Omega_m\neq0$ only when $m\in\{-F,\ldots,F-1\}$, Eq.~\eqref{eq:H} has open boundary conditions along ${\bf e}_m$, with sharp edges at $m=\pm F$. By gauge-transforming $a_{n,m}$ and $a^\dag_{n,m}$, the hopping phase $\exp(i 2\kr x)$ can be transferred to the hopping along $\ex$. Combining this  with a Fourier transformation along $\ex$, $b_{q,\, m}^{\dag}=N^{-1/2}\sum_{n=1}^{N}a_{n,\, m}^{\dag}e^{i\left(q+\gamma m\right)n}$, splits the Hamiltonian $H=\sum_{q}H_{q}$ into momentum components
\begin{align*}
H_{q}&=\sum_{m=-F}^{F}\varepsilon_{q+\gamma m}b_{q,\, m}^{\dag}b_{q,\, m}+\left(\Omega_{m}b_{q,\, m+1}^{\dag}b_{q,\, m}+{\rm H.c.}\right),
\end{align*}
where $\varepsilon_{k}=-2t\cos(k)$, $q\equiv 2\pi l/N$, and $l\in\{1,\dots,N\}$.  Figu\-re~\ref{fig:openBCspectrum} shows the resulting band structure for $F=1$.  Away from the avoided crossings, the lowest band describes the propagation of ``edge states'' localized in spin space at $m=\pm F$ (blue and red arrows): these states propagate along $\ex$ in opposite directions.  In the physical system, these give rise to a spin current $j_{s}(x)=j_{\uparrow} - j_{\downarrow}$.  When $W=2F+1 \gg 1$, these edge states become analogous to those in quantum Hall systems~\cite{SuppMat,Hugel:2013}.  The $F=9/2$ manifold of $^{40}{\rm K}$ allows experimental access to this large-$W$ limit~\cite{Hatsugai:1993}, since its 10 internal states reproduce the Hofstadter-butterfly topological band structure.

\begin{figure}
\begin{centering}
\includegraphics[width=1.05\columnwidth]{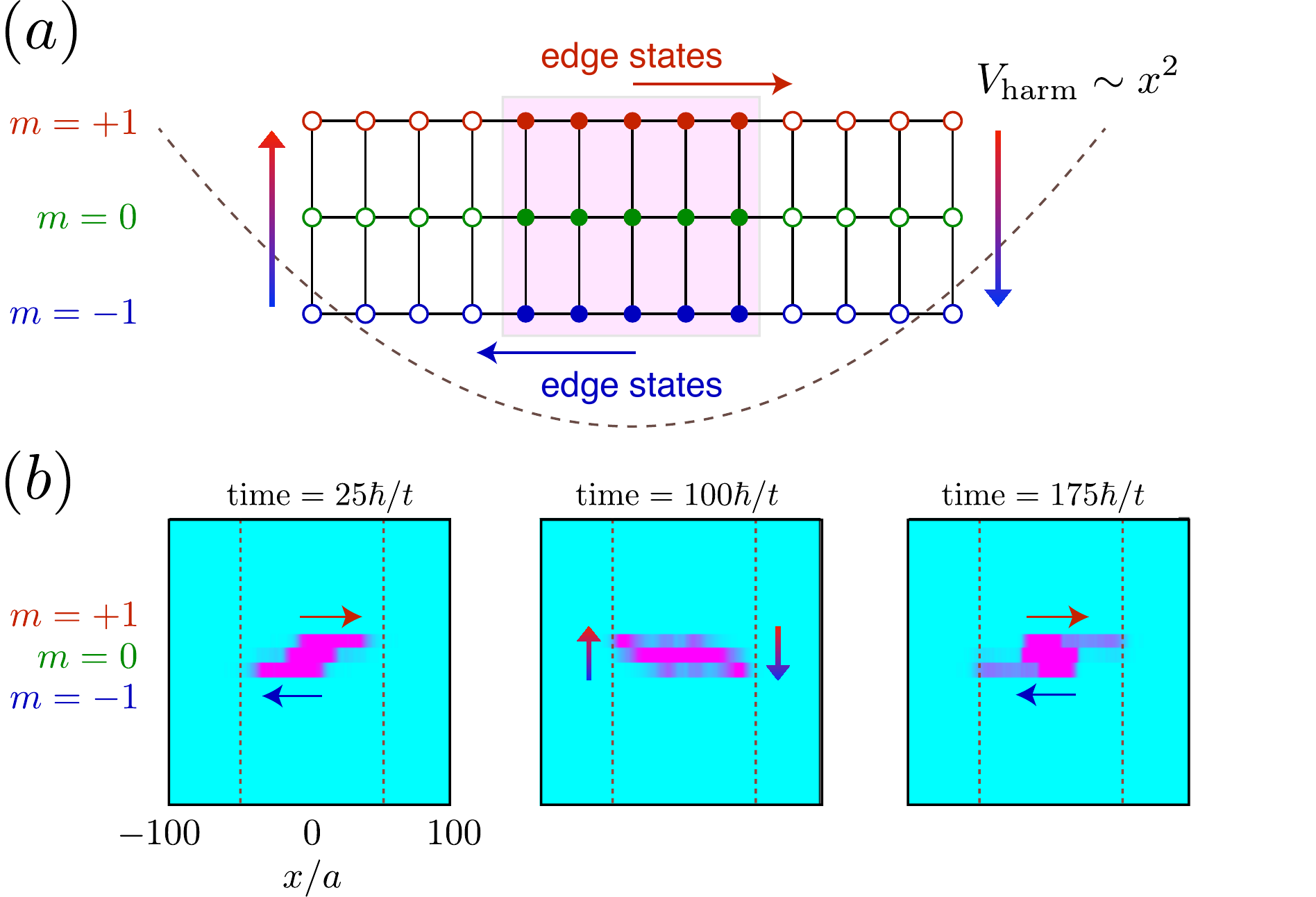}
\end{centering}
\caption{(a) Initial condition: a Fermi gas is trapped in the central region $x \in [-13a , 13a]$ and the Fermi energy is set to populate only the lowest energy band. The occupied edge states localized at $m=\pm F$ have opposite group velocities (for simplicity we sketch the ``F=1'' case). An additional harmonic potential limits the edge-states propagation, leading to chiral dynamics around the synthetic 2D lattice. (b) Dynamics after releasing the cloud into the harmonic potential, for $\Omega_0=0.5 t$, $\Phi=1/2 \pi$, $V_{\text{harm}}(x) = t (x/50 a)^2$ and $E_{\text{F}}\!=\!-1.4t$. Dashed lines represent the Fermi radius $R_{\text{F}}$ at which the edge states localized at $m\!=\!\pm F$ jump to the opposite edge $m\!=\! \mp F$.}
\label{fig:dynamics}
\end{figure}

The edge-state propagation can be directly visualized by confining a multi-component Fermi gas to a region $x \in [-L_x/2 , L_x/2]$ and by setting the Fermi energy $E_{\text{F}}$ within the Raman-induced gap (dashed line in Fig.~\ref{fig:openBCspectrum}) \cite{Note2}.  In this configuration, different types of states are initially populated: (a) edge states localized at $m=\pm F$ with opposite group velocities, and (b) bulk states delocalized in spin space with small group velocities (the central or bulk region of the lowest band is almost dispersionless for small flux $\Phi\ll1$). When the confining potential along $\ex$ is suddenly released, the edge states at $m=\pm F$ propagate along $\pm\ex$. Figure~\ref{fig:dynamics} depicts such dynamics, where we allowed tightly confined atoms (as above) to expand into a harmonic potential $V_{\text{harm}}(x)$. This potential limits the propagation of the edge states along $\ex$ and leads to chiral dynamics around the synthetic 2D lattice: when an edge state localized at $m=+ F$ reaches the Fermi radius $x=R_{\text{F}}$, it cannot backscatter because of its chiral nature, and thus, it is obliged to jump on the other edge located at $m=- F$ and counter-propagate. The edge-state dynamics of the $F=9/2$ lattice is presented in~\cite{SuppMat}.

An interesting feature of edge states is their robustness against local perturbations. To check this in the context of our proposal, we consider the effects of a spatially localized impurity on the transmission probability.
 The Hamiltonian with an impurity localized at $n=0$ is
\begin{equation}
H_{\mathrm{imp}}=H+V\,,\quad V=\sum_{m}V_{m}a^{\dag}_{0,m}a_{0,m}\,,
\label{eq:H_imp}
\end{equation}
where the zero-th order Hamiltonian $H$ is given by Eq.~\eqref{eq:H}, and $V_m$ is the interaction potential between the impurity and atoms in state $m$. The perturbation may be generated, e.g., by a tightly focused laser, or by a distinguishable atom, deeply trapped by a species selective optical lattice \cite{Massignan2006,Nishida2008,Lamporesi2010}
If the impurity scatters equally strongly with all spin components, it corresponds to an extended obstacle along ${\bf e}_m$: a ``roadblock'' in the synthetic 2D lattice. On the other hand, if the impurity interacts significantly only with a given spin component, it yields a localized perturbation in the synthetic 2D lattice. In particular, edge perturbations can be engineered by choosing an impurity that only scatters strongly the $m=F$ or $m=-F$ states. 

\begin{figure}
\begin{centering}
\includegraphics[width=.98\columnwidth]{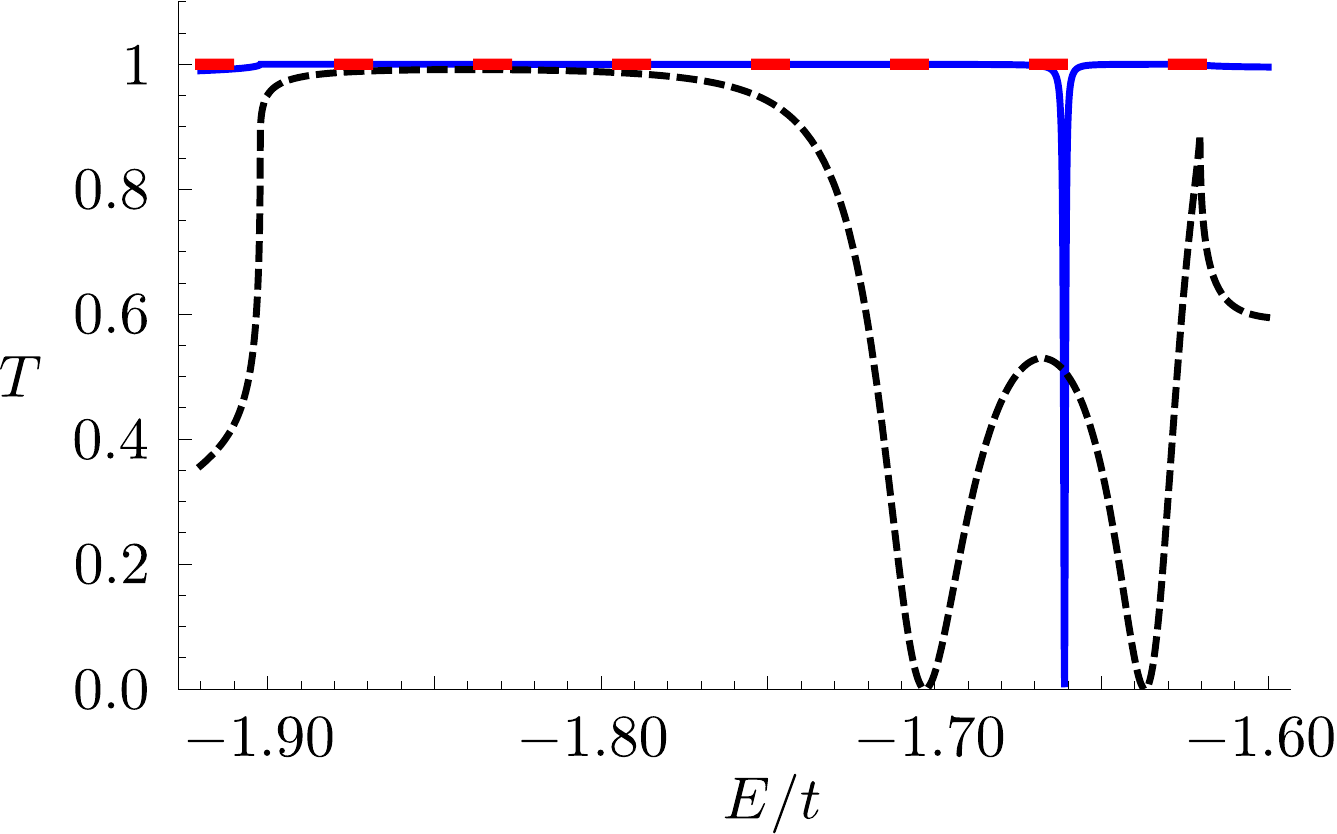}
\end{centering}
\caption{Edge-state transmission probability. Black: a spin-independent impurity.  Blue: only $m=0$ scatters.  Red: only $m=1$ scatters.  Parameters are the same as in Fig.~\ref{fig:openBCspectrum}(a) and the scattering strength is $U=-t$.}
\label{fig:transmission}
\end{figure}

For $F=1$ there are 3 dispersion branches, as shown in Fig.~\ref{fig:openBCspectrum}, so there are 9 possible scattering channels. However, here we focus to the energy range lying inside the bulk-gap (around the dashed lines in Fig.\ \ref{fig:openBCspectrum}), where there is only one available scattering channel, i.e., scattering to the opposite edge state.
The transmission probability as a function of the energy of the incident atom is calculated in~\cite{SuppMat}, and shown in Fig.~\ref{fig:transmission}. For spin-independent collisions with the impurity ($V_m=U$), the transmission probability goes to zero at two values of the energy within the gap.  
In analogy with Fano resonances~\cite{Fano1961,Satanin2005}, these zeros are associated with two quasi-bound states localized around the impurity potential due to two local parabolic minima (for $F=1$) in the upper dispersion branches. Outside of the resonant regions the transmission probability is close to 1.
On the other hand, an impurity which scatters only the $m=0$ component ($V_m=U\delta_{m,0}$) is effectively localized in the central chain of the synthetic 2D lattice. As such, it can couple resonantly two oppositely propagating edge states, leading to a single sharp minimum in the transmission probability.
Instead, an impurity which is localized at the edge of the synthetic dimension (e.g., $V_m=U\delta_{m,1}$) does not lead to a resonant behavior of the transmission probability. For such spin-dependent impurity the transmission probability is always close to 1, since the edge state can go around the impurity in the synthetic dimension.

{\it Cyclic couplings.}  In our $F=1$ example, periodic boundary conditions along ${\bf e}_m$ can be induced with an extra coupling (with a Rabi frequency $\Omega_{1}=\Omega_{\rm pbc}=\Omega_0$) from $\ket{m=1}$ to $\ket{m=-1}$ accompanied by the momentum recoil $k$ along $\ex$. The system becomes periodic only
provided the flux $\gamma$ per plaquette is rational, i.e., $\gamma= 2\pi P/Q$ with $P,Q$ co-prime integers. 
Note that the number of loops in the synthetic dimension required to have an integral number of flux quanta, i.e. periodicity, is $l/M$ where $l=LCM(M,Q)$, thus, for M=3, $Q$ or $Q/3$ loops.
 
\begin{figure}
\begin{centering}
\includegraphics[width=.98\columnwidth]{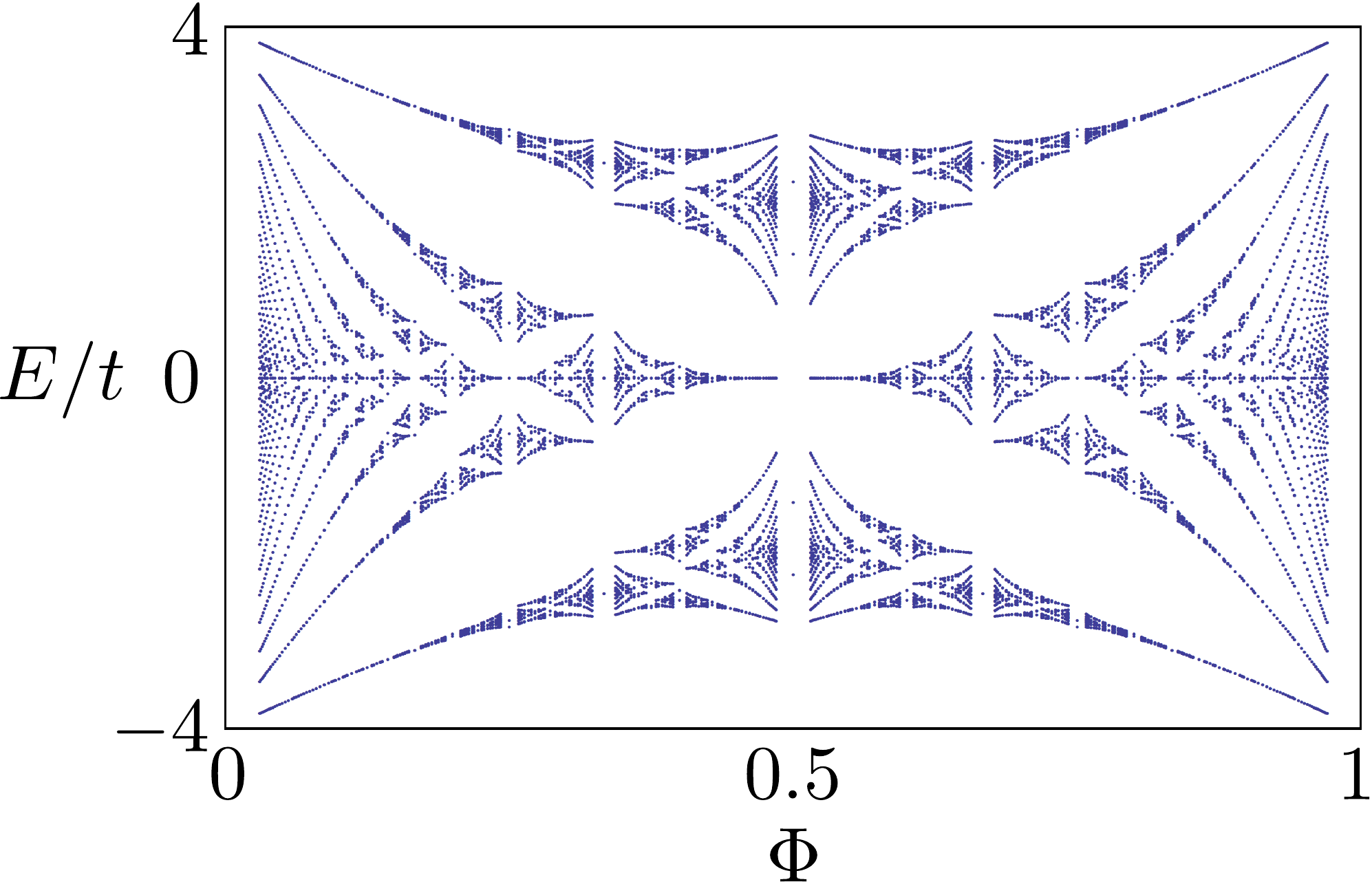}
\end{centering}
\caption{The spectrum of Eq.~(\ref{HofstadterButterflyHamiltonian}) on an infinite 1D chain, for a three-level system with closed b.c.\ has the typical {\it Hofstadter butterfly} characteristics.}
\label{fig:HofstadterButterfly}
\end{figure}

In this cyclic scheme, the system reproduces the Hofstadter problem defined in the infinite plane: its spectrum $E=E(p)$ is obtained by solving the Harper equation along 
$\ex$~\cite{Hatsugai:1993}, where $p$ is the quasi-momentum associated with the closed synthetic dimension $\ey$.  The conserved momentum along $\ey$ can only take three values: $p_j=2\pi j/3$ with $j\in\{-1,0,1\}$. Exploiting the fact that the Hamiltonian \eqref{eq:H} with closed b.c.\ is translationally invariant in the spin dimension, we perform the Fourier transform
$a_{n,m}^\dag=3^{-1/2}\sum_{j=-1}^1 e^{i 2\pi m j/3} c_{n,j}^\dag$, giving 
\begin{equation}
H=\sum_{j,n}\epsilon(2\pi j/3 +n\gamma)c^{\dag}_{n,j} c_{n,j} - (tc^{\dag}_{n+1,j} c_{n,j} + {\rm H.c.}),
\label{HofstadterButterflyHamiltonian}
\end{equation}
and $\epsilon(k)=-2\Omega_0 \cos(k)$. Its spectrum is plotted in Fig.~\ref{fig:HofstadterButterfly}.
There are  $l$ points  in each band associated with the rational flux $\gamma$: enough to be visible.  For our finite chain of length $N$, the infinite-chain result will be accurate only for $Q\ll N$, while for $Q$ approaching $N$ the system is far from periodic in $Q$ and the butterfly gets blurred.

{\it Interactions.}  We wish to consider the effects of repulsive interactions. We focus here on the case where the interactions are SU(W)-invariant (this amounts to negleglecting the spin-dependent contribution to the interaction; a very good approximation for $F=1$ $\Rb87$).  In our lattice, the resulting interaction Hamiltonian
\begin{align*}
H_{\rm int} &=\frac{\mathcal{U}}{2}\sum_{n} \mathcal{N}_n (\mathcal{N}_n -1)\,, \qquad \mathcal{N}_n \equiv \sum_m  a_{n,\, m}^{\dagger}a_{n,\, m},
\end{align*}
is local along $\ex$, but infinite in range along ${\bf e}_m$. 
We exploit the SU(W)-invariance of $H_{\rm int}$ by adopting the Fock basis $c_{n,j}$ in which the hopping along ${\bf e}_m$ is diagonal, as in Eq.~\eqref{HofstadterButterflyHamiltonian} (a similar basis exists for open boundary conditions in the synthetic dimension). Let us denote its eigenvalues by $\epsilon_{n,j}$. It follows that we can minimize the energy for fixed $\langle H_{\rm int} \rangle$ by populating only the states associated to $c_{n,j_n}$ with lowest  $\epsilon_{n,j_n}$, as this minimizes the kinetic term $\langle H \rangle$.

Two cases are possible: i) $j_n$ is unique, i.e. the local ground state is not degenerate; ii) $\epsilon_{j,n}$ is minimal for two of the three possible values of $j$. The latter case can occur only for closed b.c. in the synthetic dimension and for rational values of the flux $\gamma/(2\pi)=P/Q$. In presence of open b.c., it is indeed easy to show that the eigenvalues are always independent of $\gamma$ (and as such as $n$), and never degenerate.
In case i), the ground state can be mapped to the one of a 1D uniform Bose-Hubbard chain. In case ii) instead, the 1D Hubbard chain will possess a primitive cell containg $Q$ consecutive lattice points, as well known from the non-interacting Hofstadter problem. Interactions which are non-SU(N)-invariant lead to considerably more complicated situations, with the ground state possessing a complex, fully 2D character.
{\it Conclusions.} Our proposal for creating strong synthetic gauge fields using a synthetic 2D lattice is well suited to directly observe chiral edge-states dynamics, by using spin-sensitive detection of the different edge modes.  This platform also allows to test the edge states' robustness against impurities.  To detect the full spectrum, interaction effects must be minimized, for example using a fermionic band insulator or a dilute thermal Bose gas.  
The spectrum may also be probed by transport measurements: wavepackets of atoms with narrow energy dispersion  can be prepared and brought into the lattice using a waveguide, and their transmission through the region of effective magnetic field observed~\cite{Lauber2011,Cheiney2013}.

\acknowledgments{
We acknowledge enlightening discussions with E. Anisimovas, F. Chevy, J. Dalibard, L. Fallani, F. Gerbier and C. Salomon and support from FRS-FNRS (Belgium), ERC AdG QUAGATUA, EU IP SIQS, Spanish MINCIN (FIS2008-00784 TOQATA), ESF POLATOM network, the European Social Fund under the Global Grant measure.  IBS acknowledges the support of the ARO with funding from DARPA's OLE program and the Atomtronics-MURI; and the NSF through the PFC at JQI.
}

\bibliographystyle{prsty}

\vspace{5cm}

\section{Supplemental material}

\subsection{Edge states in thin stripes: Hofstadter square lattice vs Hofstadter ladder}
\label{appendix:hofstadter}

In the main text, we have concentrated on the spectra and edge-state dynamics for spin 1 atoms ($F=1$).  In that case a synthetic 2D lattice is constituted of $N \times 3$ lattice sites, where $3=W=2 F+1$ is the number of sites along the synthetic (spin) direction and $N$ is the number of sites along the spatial direction $x$ (see Figs. 1 -- 3 in the main text). Such a lattice has natural open boundaries along the spin direction at $y=\pm F a$ (where $a$ is the lattice spacing), while $N$ can be arbitrarily large. In this Appendix, we illustrate how the edge-state properties discussed in the main text can be related to the topological band structure and chiral edge states of the standard Hofstadter square lattice~\cite{Hatsugai:1993}, namely, a square lattice of $N \times W$ sites, with $N,W \gg 1$, subjected to a uniform magnetic flux $\Phi$ per plaquette. The number of lattice sites along the $y$ direction is denoted $W$, so as to refer to the width of the stripe.

\begin{figure}
\begin{center}
\includegraphics[width=3.5in]{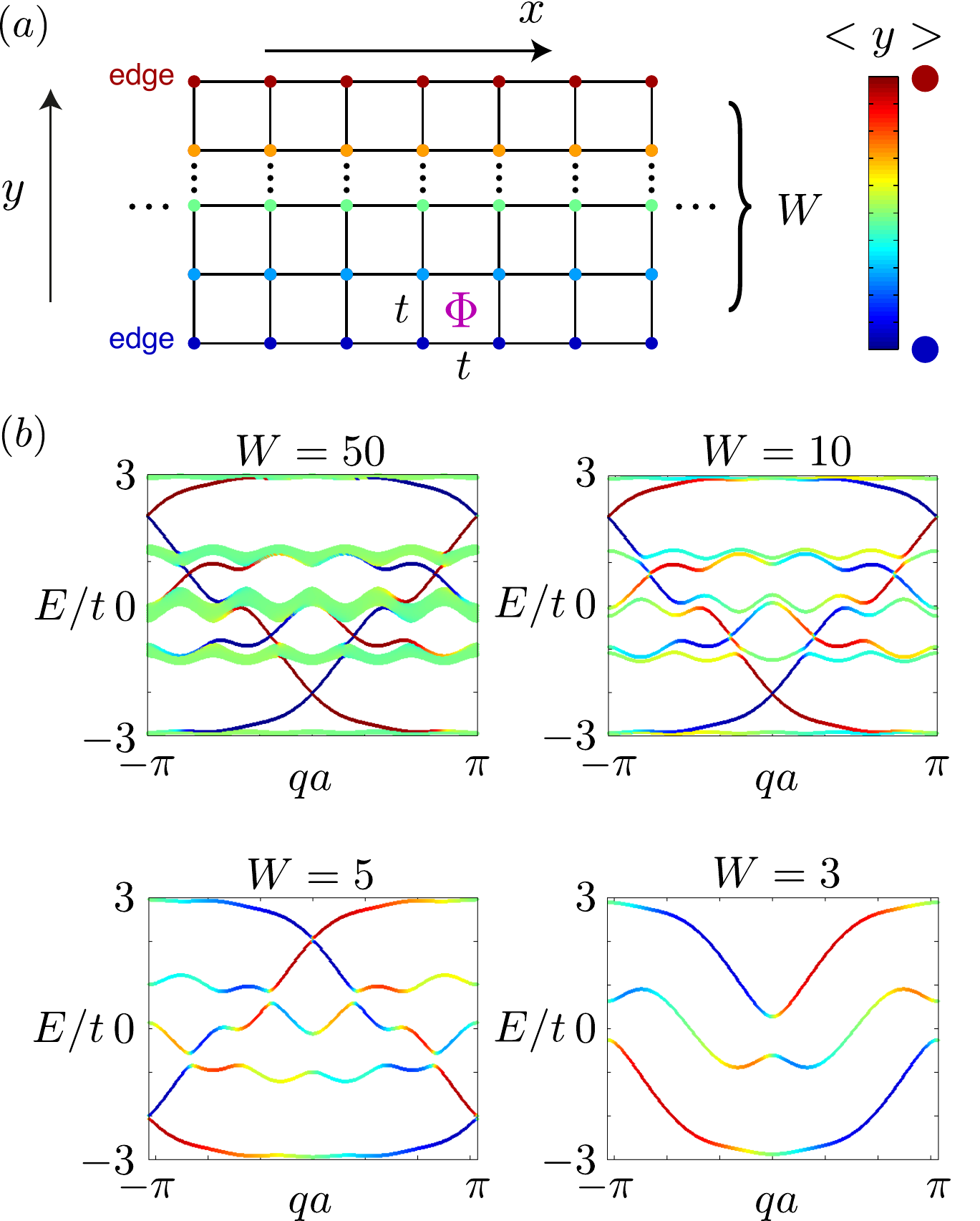}
\end{center}
\caption{(a) Hofstadter model on a stripe of width $W$, and definition of the color code: dark blue (resp. red) dots correspond to states localized at the bottom (resp. top) edge of the system, whereas green-yellow dots correspond to bulk states. (b) Energy spectrum $E=E(q)$ of the Hofstadter model with the flux $\Phi=1/5$, for different stripe widths $W$. Here, the modulus of the hopping amplitude is taken equal to $t$ along both directions, and $q$ denotes the quasi-momentum. The double-ladder configuration used in the main text corresponds to $W=3$ (i.e., $F=1$ and $\Omega_0=t$).}
\label{FIGSuper}
\end{figure}

To do so, we consider an extrapolation between the Hofstadter lattice (size $N \times W$) and the thin stripe considered in the main text (size $N \times 3$), by progressively reducing the number of lattice sites along the $y$ direction $W$, while applying periodic boundary conditions along the $x$ direction, see Fig. \ref{FIGSuper} (a). The first spectrum shown in Fig. \ref{FIGSuper} (b), obtained for $W=50$, shows the usual band structure of the Hofstadter model, where a clear distinction between the bulk bands and the edge states dispersions is observed. To highlight this edge/bulk picture, we simultaneously represent the energies $E=E(q)$ together with the mean position $\langle y \rangle$ of the eigenstates along the spin direction, see the color code in Fig. \ref{FIGSuper} (a). The many bulk states progressively disappear, as the number of inequivalent lattice sites is reduced to $W=5$, while the dispersion branches of the edge states are only slightly modified. In fact, for $\Phi=p/q \in \mathbb{Q}$, the edge-state branches remain remarkably robust for $W \rightarrow q$. When $W$ is further reduced such that $W<q$, the edge-state branches are altered, but they retain their general characteristics: in the thin stripe (``double-ladder") limit $W=3$ considered in the main text, the lowest energy band describes edges states localized on opposite edges (at $y=\pm a$) of the double-ladder, propagating in opposite directions. Therefore, we can conclude that the edge-state structure present in the double-ladder lattice ($W=3$) is reminiscent of the chiral (topological) edge states present in the standard Hofstadter square lattice (see also Ref.~\onlinecite{Hugel:2013} for a detailed study of the Hofstadter ladder with $W=2$ corresponding to $F=1/2$). 

\subsection{The $F=9/2$ case}
\label{appendix:potassium}

\begin{figure}
\begin{center}
\includegraphics[width=3.5in]{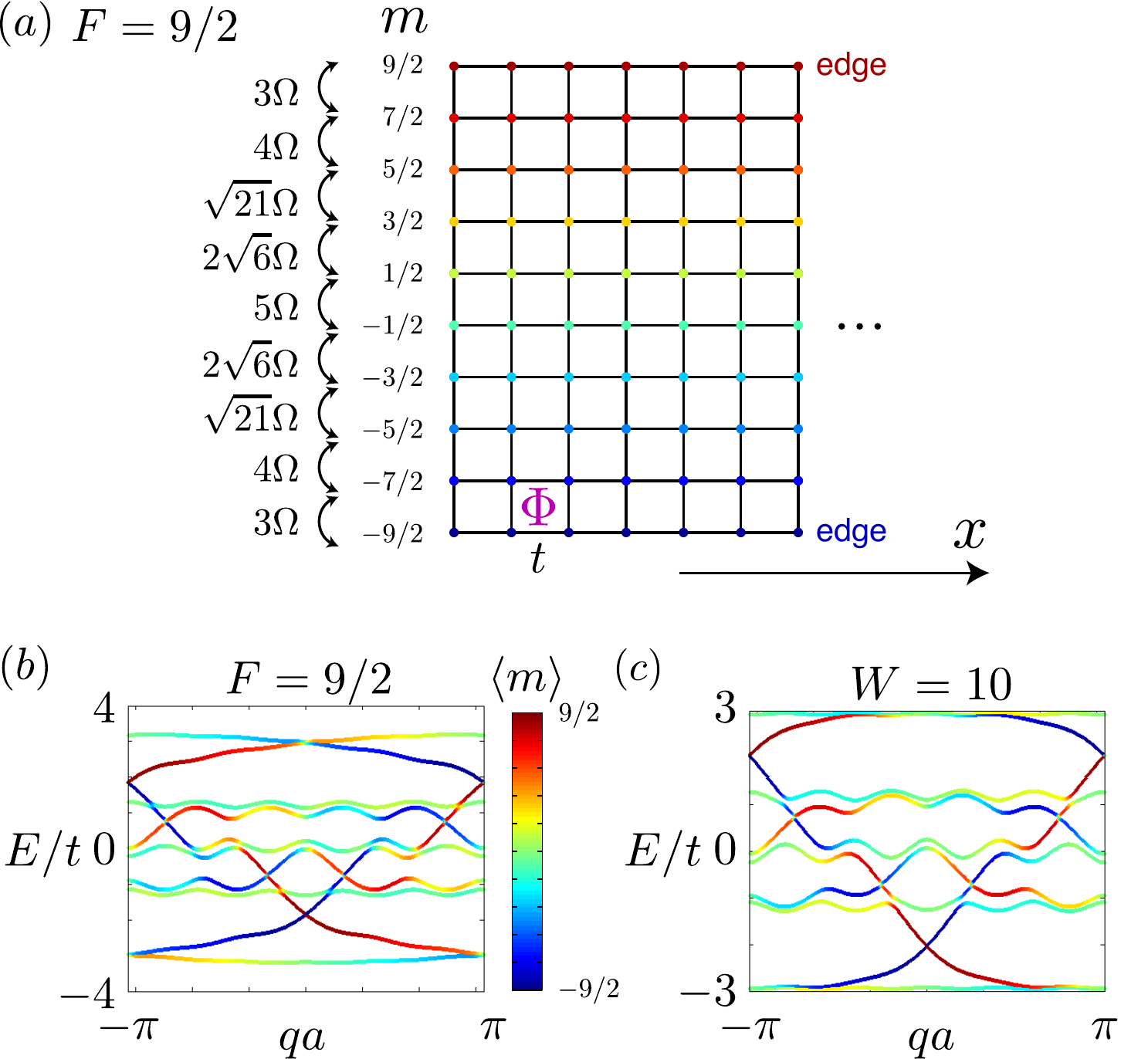}
\end{center}
\caption{(a) Synthetic lattice for $F=9/2$ atoms. The hopping amplitude $t$ along the $x$ (spatial) direction is constant, while the hopping amplitude along the $y$ (spin) direction, $\Omega g_{F,m}$, is given by Eq. \eqref{spin_hop_eq}. (b) The energy spectrum for the $F=9/2$ synthetic lattice, setting $\Phi=1/5$ and $\Omega = t / \langle g_{F,m} \rangle = 0.24 t$.
 (c) The energy spectrum for the homogenous Hofstadter lattice with $W=10$ lattice sites along the $y$ direction and $\Phi=1/5$, see also Fig. \ref{FIGSuper}b. Note that the edge states are more spatially localized in the homogeneous case [(c)] than in the inhomogeneous synthetic lattice [(b)].}
\label{FIGPotassium}
\end{figure}

In the main text, we focused on the study of the $F=1$ case, which is widely investigated in current cold-atom experiments~\cite{Lin2009a,Lin2009b}. This leads to the double-ladder lattice, whose connection with the standard Hofstadter model has been described in the previous Section of this Supplementary material. However, it would be desirable to engineer a synthetic 2D lattice with more internal states to make this connection even more visible. For example, considering the ground-state manifold of $^{40}$K, where $F=9/2$, would allow to engineer a lattice of size $N \times 10$, which according to Fig. \ref{FIGSuper} (b) would clearly display the topological band structure of the Hofstadter model. We note that using other atomic species (such as $^{173}$Yb) could also lead to similar configurations with $W > 5$, both for bosonic and fermionic systems. \\
One important aspect of the present proposal is the fact that for $F>1$ the magnitude of hopping along the $y$ (spin) direction is not constant. Indeed, the hopping from a lattice site $m$ to a lattice site $m+1$ is given by the frequency 
\begin{equation}
t_{m \rightarrow m+1} = \Omega g_{F,m} = \Omega \sqrt{F(F+1) - m(m+1)},
\label{spin_hop_eq}
\end{equation}
where we remind that $m=m$ refers to the internal states of the atom and $F$ is the total angular momentum. This inhomogenous hopping, shown in Fig. \ref{FIGPotassium} (a) for $F=9/2$, is not present in the standard Hofstadter model, where the tight-binding hopping amplitude $t$ is constant. To illustrate this effect, we show the band structure of a synthetic lattice engineered with $F=9/2$ atoms (Fig. \ref{FIGPotassium} (b)), and we compare it with the band structure of the homogenous Hofstadter model with $W=10$ (Fig. \ref{FIGPotassium} (c)). We observe that the bulk/edge band structure is well conserved, when choosing $\Omega = t / \langle g_{F,m} \rangle$, where $\langle g_{F,m} \rangle =\sum_m g_{F,m}/2 F$. However, we note that the states corresponding to the edge-state dispersions are no longer perfectly localized at the edges: close to the lowest bulk band, there are dispersive states with $\vert \langle m \rangle \vert < 9/2 $. We also note that the states with the highest velocity $v \!\sim\! \partial_q E$ are those that are the most localized at the edges.

\begin{figure}
\begin{center}
\includegraphics[width=3.55in]{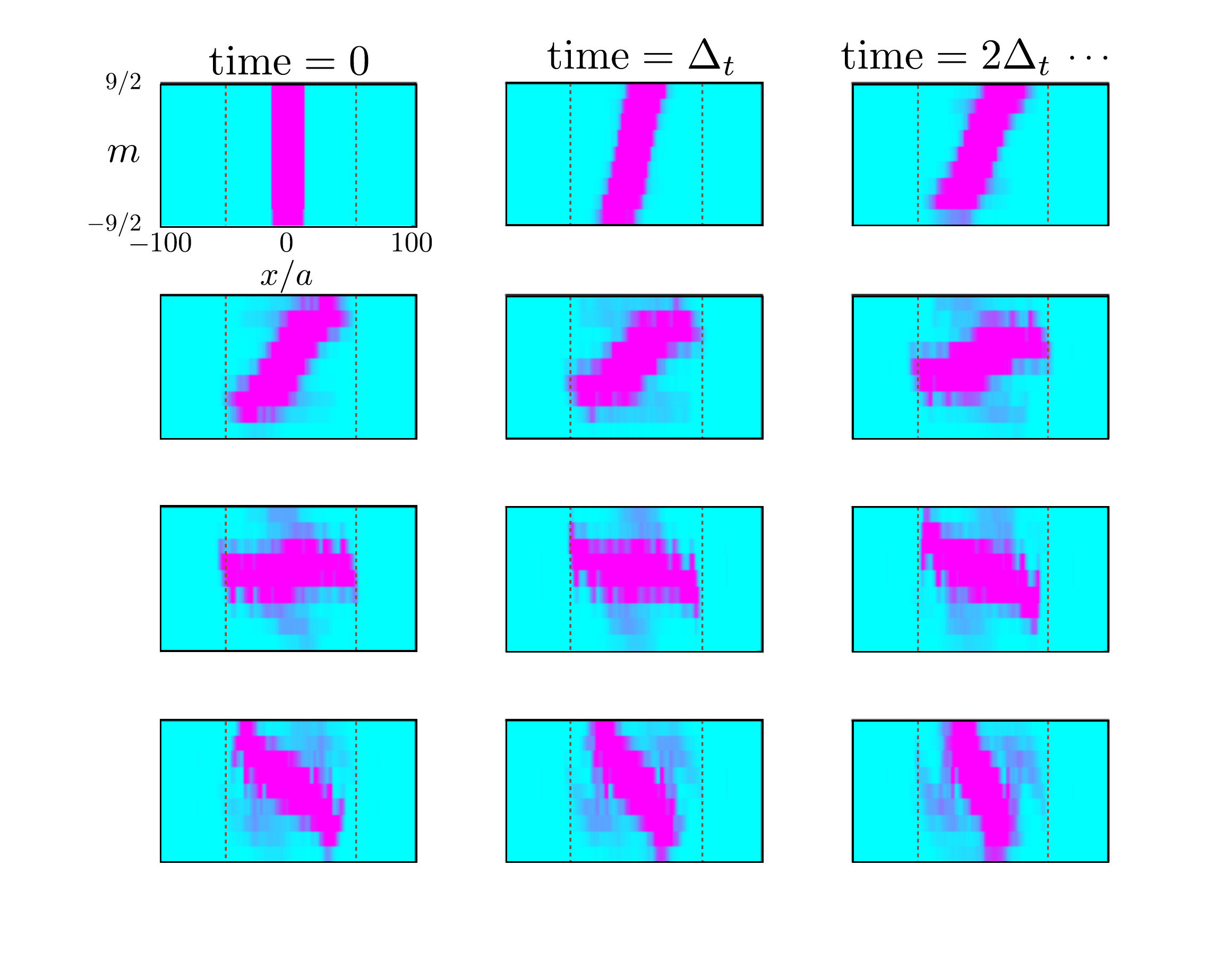}
\end{center}
\caption{Edge-states dynamics for a fermionic system with $F=9/2$ atoms (e.g. $^{40}K$): the Fermi gas is trapped in the central region $x \in [-13a , 13a]$ and the Fermi energy is set such as to populate only the lowest energy band. The populated ``edge" states localized at $m=\pm F$ have opposite group velocities. An additional harmonic potential limits the edge-states propagation, leading to chiral dynamics around the synthetic 2D lattice. The parameters are $\Omega = t / \langle g_{F,m} \rangle = 0.24 t$, $\Phi=1/5$, $V_{\text{harm}}(x) = t (x/50 a)^2$ and $E_{\text{F}}\!=\!-2t$. Dashed lines represent the Fermi radius $R_{\text{F}}$ at which the edge states localized at $m\!=\!\pm F$ jump unto the opposite edge $m\!=\! \mp F$. The time steps are $\Delta_t=37.5 \hbar/J$.}
\label{FIGDymPotassium}
\end{figure}

In Fig. \ref{FIGDymPotassium}, we show the edge-state dynamics for a fermionic system with $F=9/2$ atoms (e.g. $^{40}K$), confined by a harmonic potential $V_{\text{harm}} (x) = t (x/50a)^2$. We clearly observe a chiral motion in the 2D synthetic lattice, which is due to the populated edge states lying within the lowest bulk gap (Fig. \ref{FIGPotassium} (b)). As already described above, these edge states are not perfectly localized at $m=\pm 9/2$, due to the inhomogeneity of the hopping along the spin direction. As a result, the dynamics show the rotation of the cloud in the 2D lattice, instead of a clear edge-state motion. 

\subsection{Scattering on a localized impurity}

\subsubsection{Formulation}

Our aim here is to calculate the transmission probability for an atom in the 1D physical lattice affected by an impurity localized at $n=0$ and thus described by the Hamiltonian 
\begin{equation}
H_{\mathrm{imp}}=H+V\,,\quad V=\sum_{m,m'}V_{m,m'}a^{\dag}_{0,m}a_{0,m'}\,,
\label{eq:H_imp_app}
\end{equation}
where $H$ is an unperturbed Hamiltonian for the 1D array of atoms is given by Eq.(\ref{eq:H}) of the main text, and $m$ refers to the spin levels representing a synthetic degree of freedom.  

We shall make use of the Green's operator $G=[E-H_{\mathrm{imp}}+i0^{+}]^{-1}$ of the full Hamiltonian $H_{\mathrm{imp}}$. 
The Green's operator of the complete system will be expressed in terms of the Green's operator $G_0=[E-H+i0^{+}]^{-1}$ of the unperturbed system using the Dyson equation~\cite{Economou2006} $G=G_0+G_0VG$.
On the other hand, the zero-order Green's operator $G_0$ will be presented via the eigenfunctions and eigen-energies of the unperturbed Hamiltonian $H$. Having the complete Green's operator $G$ we will determine the scattering T-matrix $T=V+VGV$ from which the transmission probabilities will be calculated. 

\subsubsection{Spectrum of the Hamiltonian without impurity}

Applying a gauge transformation $\tilde{a}_{n,m}=a_{n,m}e^{-i\gamma nm}$
we transfer the phases featured in the hopping elements to the hopping in the physical direction
in the Hamiltonian $H$ defined by Eq.~(3) in the main text, giving:
\begin{equation}
H=\sum_{n,m}\left(-te^{-i\gamma m}\tilde{a}_{n+1,m}^{\dag}+\Omega_{m-1}\tilde{a}_{n,m-1}^{\dag}\right)\tilde{a}_{n,m}+\mathrm{h.c.}\,.
\label{eq:H-suppl-1}
\end{equation}
From now on we will express all energies in the units of the hopping
integral $t$; therefore, we will set $t=1$. The atomic center-of
mass wave function satisfies the Schr\"odinger equation
\begin{equation}
H\Psi=E\Psi \,.
\label{eq:schroed}
\end{equation}
We search for the eigenvectors of the Hamiltonian~\eqref{eq:H-suppl-1} in
the form of plane waves (Bloch states) by taking the probability amplitudes
to fi{}nd an atom in the site $n,m$ as
\begin{equation}
\Psi_{m}(n)=\chi_{q,m}e^{iqn}\,.
\end{equation}
We will interpret the index $m$ as a row number and consider $\Psi$
and $\chi_{q}$ as columns. Equation (\ref{eq:schroed}) yields the
following eigenvalue equations
\[
H_{q}\chi_{q}=E_{q}\chi_{q}\,.
\]
Here $H_{q}$ is $(2F+1)\times(2F+1)$ matrix with the diagonal matrix
elements $(H_{q})_{m,m}=-2\cos(q+\gamma m)$ and nonzero non-diagonal
elements $(H_{q})_{m,m'}=\Omega_{m}\delta_{m',m+1}$ and $(H_{q})_{m,m'}=\Omega_{m-1}\delta_{m',m-1}$.
In particular, when $F=1$ the matrix $H_{q}$ reduces to
\begin{equation}
H_{q}=\left(\begin{array}{ccc}
-2\cos(q-\gamma) & \Omega & 0\\
\Omega & -2\cos(q) & \Omega\\
0 & \Omega & -2\cos(q+\gamma)
\end{array}\right)\,.
\end{equation}
By solving an eigenvalue problem we get a set of $2F+1$ algebraic equations. It has has $2F+1$ solutions 
to be labelled with an index $\nu$.

\subsubsection{Green's function of the system without impurity}

Given the eigenfunctions $\Psi_{q,s}(n)$, the general expression
for the retarded zero-order Green's function is
\begin{equation}
G_{0}(n,n';E)=\sum_{\nu=1}^{2F+1}\int_{-\pi}^{\pi}\frac{\Psi_{q,\nu}(n)\Psi_{q,\nu}^{*}(n')}{E-E_{q,\nu}+i\eta}dq\,,
\label{eq:green-general}
\end{equation}
where $\eta\rightarrow+0$. Zeros in the denominator can be obtained
from the eigen-energy equation
\begin{equation}
\det[E-H_{q}]=0\,,
\end{equation}
which generally has $2F+1$ solutions. 
For each eigen-energy $E$ and wave vector $q_{\nu}$, the analytical expressions
for the eigenvectors $\chi_{q_{\nu},\nu}$ can be obtained from the
equation $[H_{q}-E]\chi_{q_{\nu},\nu}=0$ by setting the first element
of $\chi_{q_{\nu},\nu}$ to unity and dropping one of the resulting
equations. Using Eq.~(\ref{eq:green-general}) and performing the
integration we obtain the retarded zero-order  Green's function
\begin{equation}
G_{0}(n,n';E)=-i\sum_{\nu}\frac{1}{v_{\nu}}\begin{cases}
\chi_{q_{\nu},\nu}\chi_{q_{\nu},\nu}^{T}e^{iq_{\nu}(n-n')}\,, & n>n',\\
\chi_{-q_{\nu},\nu}\chi_{-q_{\nu},\nu}^{T}e^{-iq_{\nu}(n-n')}\,, & n<n',
\end{cases}\label{eq:green0}
\end{equation}
Here
\begin{equation}
v_{\nu}\equiv\left.\frac{\partial}{\partial q}E_{q,\nu}\right|_{q=q_{\nu}}\label{eq:velocity-1}
\end{equation}
is the group velocity. It can be calculated from the equation
\begin{equation}
v_{\nu}=-\left.\frac{\frac{\partial}{\partial q}\det[E-H_{q}]}{\frac{\partial}{\partial E}\det[E-H_{q}]}\right|_{q=q_{\nu}}\,.
\label{eq:velocity-2}
\end{equation}
Note that we do not have complex conjugation in Eq.~(\ref{eq:green0})
since for real wave vectors $q_{\nu}$ the colums $\chi_{q_{\nu},\nu}$
are real. This is because the Hamiltonian $H_{q}$ has real matrix
elements.

\subsubsection{Green's function for the system with localized impurity}

Combining the Dyson equation $G=G_0+G_0VG$ with Eq.~\eqref{eq:H_imp} for $V$, one has
\begin{eqnarray}
G(n,n') & = & G_{0}(n,n')+\sum_{n''}G_{0}(n,n'')V\delta_{n'',0}G(n'',n')\nonumber \\
 & = & G_{0}(n,n')+G_{0}(n,0)VG(0,n')\,.
 \label{eq:dyson}
\end{eqnarray}
Taking $n=0$ in Eq.~(\ref{eq:dyson}) we get
\begin{equation}
G(0,n')=G_{0}(0,n')+G_{0}(0,0)VG(0,n')\,.
\end{equation}
From here we obtain
\begin{equation}
G(0,n')=[1-G_{0}(0,0)V]^{-1}G_{0}(0,n')\,.
\label{eq:green-l0}
\end{equation}
Substituting Eq.~(\ref{eq:green-l0}) back into Eq.~(\ref{eq:dyson})
we get the required expression for the Green's function
\begin{equation}
G(n,n')=G_{0}(n,n')+G_{0}(n,0)V[1-G_{0}(0,0)V]^{-1}G_{0}(0,n')\,.
\label{eq:green-full}
\end{equation}

\subsubsection{Transmission probabilities}

The scattering is described by $T$ matrix
\begin{equation}
T=V+VGV\,.
\end{equation}
Using Eq.~(\ref{eq:green-full}), $T$ matrix reads
\begin{equation}
T(n,n')=V[1-G^{(0)}(0,0)V]^{-1}\delta_{n,0}\delta_{n',0}\,.
\label{eq:t-matrix}
\end{equation}
\begin{widetext}
For transmitted waves the matrix element of the scattering matrix is
\begin{equation}
S_{\nu,\nu'}^{t}=\delta_{\nu,\nu'}-i\frac{1}{\sqrt{v_{\nu}v_{\nu'}}}\sum_{n,n'}\chi_{q_{\nu},\nu}^{\dag}e^{-iq_{\nu}n}T(n,n')\chi_{q_{\nu'},\nu'}e^{iq_{\nu'}n'}\,.
\end{equation}
Using Eq.~(\ref{eq:t-matrix}) we obtain
\begin{equation}
S_{\nu,\nu'}^{t}=\delta_{\nu,\nu'}-\sqrt{\frac{v_{\nu}}{v_{\nu'}}}i\frac{1}{v_{\nu}}\chi_{q_{\nu},\nu}^{\dag}V\left[1+i\sum_{\nu''}\frac{1}{v_{\nu''}}\chi_{q_{\nu''},\nu''}\chi_{q_{\nu''},\nu''}^{\dag}V\right]^{-1}\chi_{q_{\nu'},\nu'}\,.
\label{eq:st}
\end{equation}
\end{widetext}
Transmission probability from the propagating mode $\nu'$ to the
mode $\nu$ is
\begin{equation}
T_{\nu,\nu'}=|S_{\nu,\nu'}^{t}|^{2}\,.
\end{equation}

These equations are used in calculating the transmission probabilities in the main text.

\end{document}